\documentclass{appolb}
\usepackage{epsfig}

\begin{document}

\title{Diffractive Physics Results at CDF
\thanks{Presented at ``XXXIII International Symposium
on Multiparticle Dynamics, September 5-11, 2003, Krak\'ow, Poland''.}
}
\author{Michele Gallinaro \footnote{Representing the CDF collaboration.} 
\address{The Rockefeller University \\ 1230 York Avenue, New York, NY 10021, USA \\ Email: michgall@fnal.gov}
}
\maketitle
\begin{abstract}

Forward detectors are described together with the first physics results from Run~II.
Using new data and dedicated diffractive triggers, a measurement of single diffractive dijet production rate, 
with particular focus on the diffractive structure function of the antiproton, is discussed.
Upper limits on the exclusive dijet and $\chi^0_c$ production cross sections are also presented.

\end{abstract}

\section{Introduction}

The signature of a typical diffractive event in $p\overline{p}$ collisions is characterized by a 
leading proton or antiproton and/or a region at large pseudorapidity with no particles, also known as {\it rapidity gap}.
Hard diffraction processes are hadronic interactions that incorporate a high transverse momentum partonic scattering, 
while carrying the characteristic signature of a diffractive event.

\begin{figure}[tp]
\epsfxsize=1.0\textwidth
\centerline{\epsfig{figure=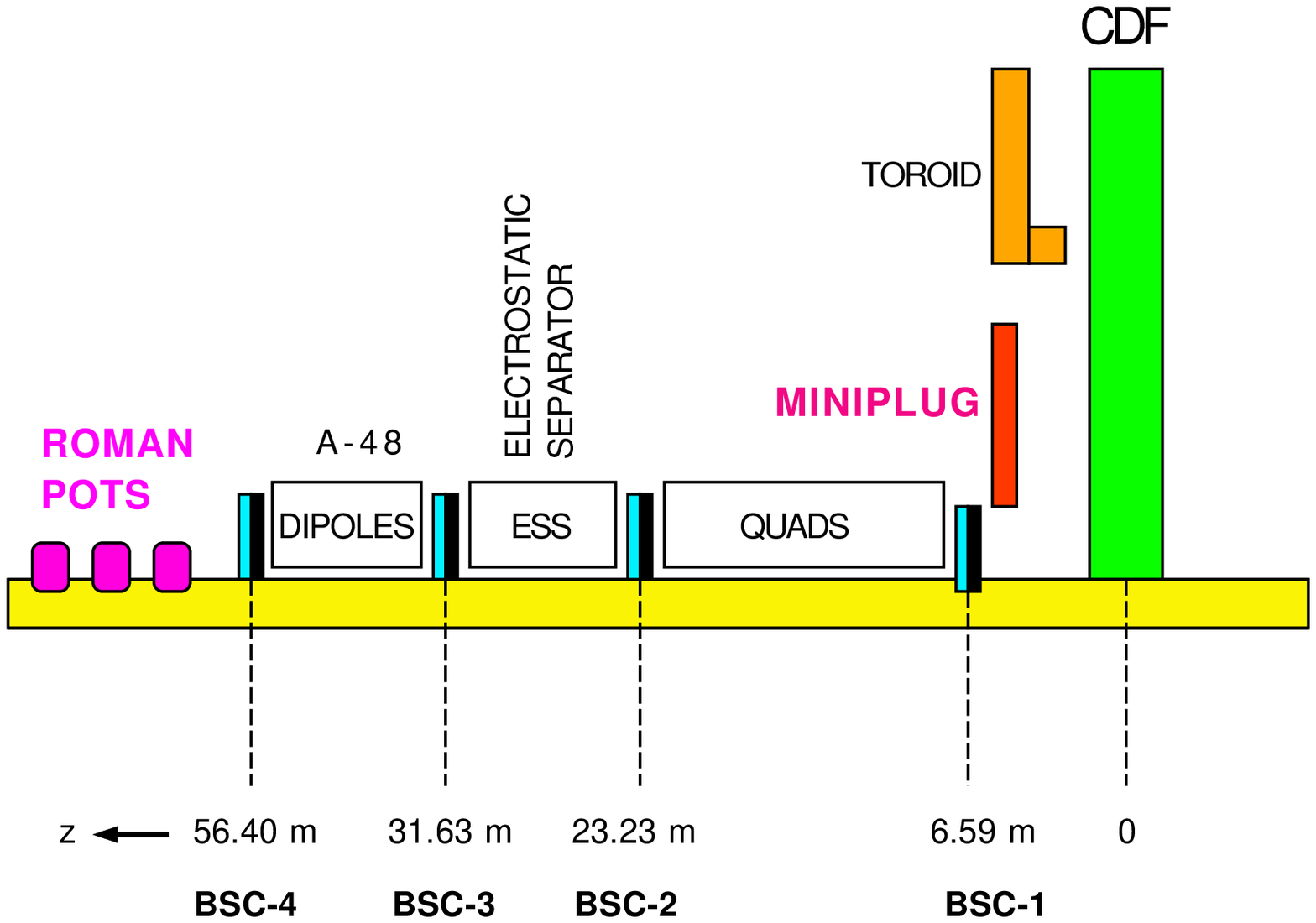,width=0.49\hsize} \hspace*{0.3cm}
	    \epsfig{figure=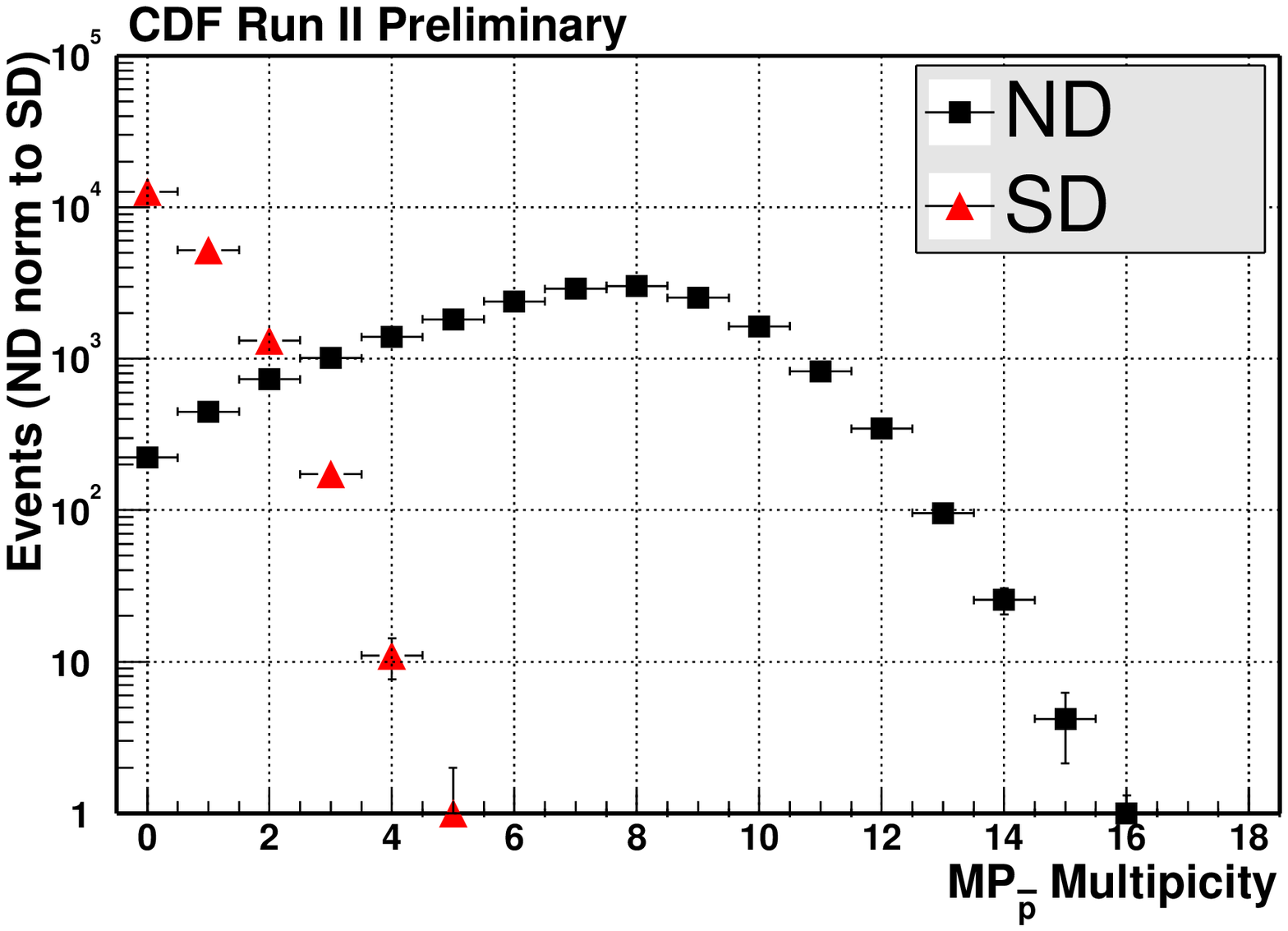,width=0.49\hsize}}
\caption{\label{diffractive_diagrams}
{\em Left}: Forward detectors along the $p\overline{p}$ direction on the west side of the CDF central detector (not to scale);
{\em Right}: Hit multiplicity in the MiniPlug calorimeters in SD and ND events.
}
\end{figure}

CDF has reported a number of diffractive studies using Run~I data~\cite{dino}.
In Run~II these measurements can be extended with larger data samples, new triggers, and improved detectors~\cite{lishep}.
CDF improved the particle coverage in the forward direction with the installation of new 
{\it Beam Shower Counters} (BSCs) covering $5.5<|\eta|<7.5$ and two forward {\it MiniPlug} (MP) calorimeters 
covering $3.5<|\eta|<5.1$ (Fig.~\ref{diffractive_diagrams}, left), which 
provide a clean separation between diffractive and non-diffractive events (Fig.~\ref{diffractive_diagrams}, right). 
The {\it Roman Pot} (RP) fiber tracker designed to detect leading antiprotons was reinstalled as in Run~I. 

First Run~II measurements include single diffractive dijet production and search for exclusive production of
dijet and $\chi^0_c$ final states.

\section{Diffractive Dijets}
Diffractive dijet events are characterized by the presence of two jets resulting 
from a hard scattering and a leading antiproton which escapes the collision intact,
only losing a small momentum fraction $\xi_{\overline{p}}$ to the Pomeron.
Events can be described in terms of a Pomeron emitted from the anti-proton and
scattering with a parton from the proton.
The gluon and quark content of the interacting partons can be investigated by comparing 
single diffractive (SD) and non-diffractive (ND) events.
The ratio of diffractive to non-diffractive dijet production rates is proportional to the ratio of the
corresponding structure functions and can be studied as a function of the Bjorken
scaling variable $x_{Bj}=\beta\cdot\xi_{\overline{p}}$ of the struck parton in the antiproton, 
where $\beta$ corresponds to the Pomeron momentum fraction carried by the parton.
For each event, $x_{Bj}$ is evaluated from the $E_T$ and $\eta$ of the jets using the equation

$$x_{Bj}=\frac{1}{\sqrt{s}}\sum_{i=1}^nE_T^ie^{-\eta^i}$$

In Run~I, CDF measured the ratio of SD to ND dijet production rates using the RP spectrometer to detect leading antiprotons. 
The CDF result~\cite{diff_dijet_rp} is suppressed by a factor of $\sim10$ relative 
to predictions from HERA data, indicating a breakdown of
conventional factorization between HERA and the Tevatron.
Correct predictions can be obtained by scaling the rapidity gap probability 
distribution of the diffractive structure function to the total integrated gap probability~\cite{dino2}.

\begin{figure}[tp]
\epsfxsize=1.0\textwidth
\centerline{\epsfig{figure=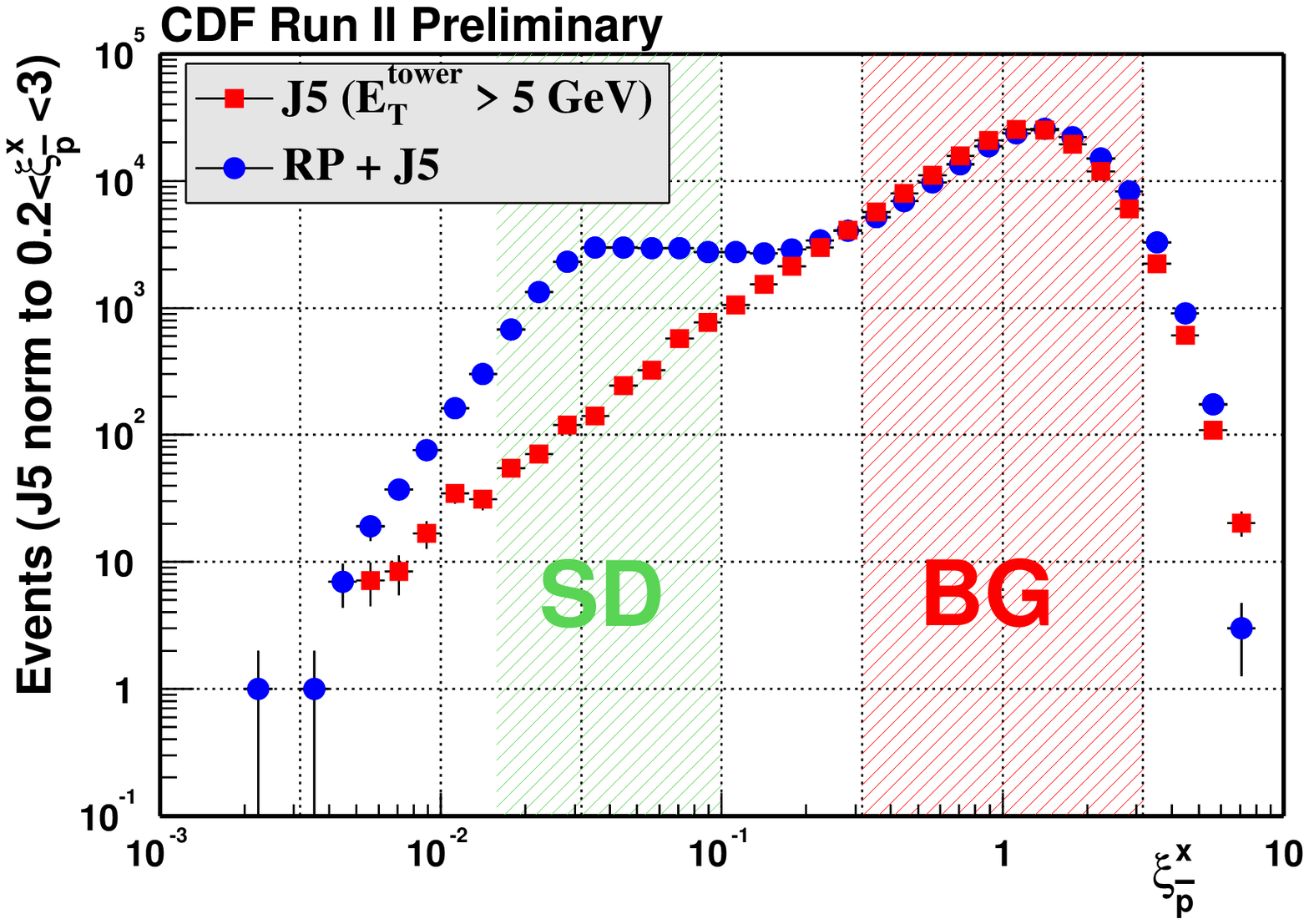,width=0.55\hsize}
	    \epsfig{figure=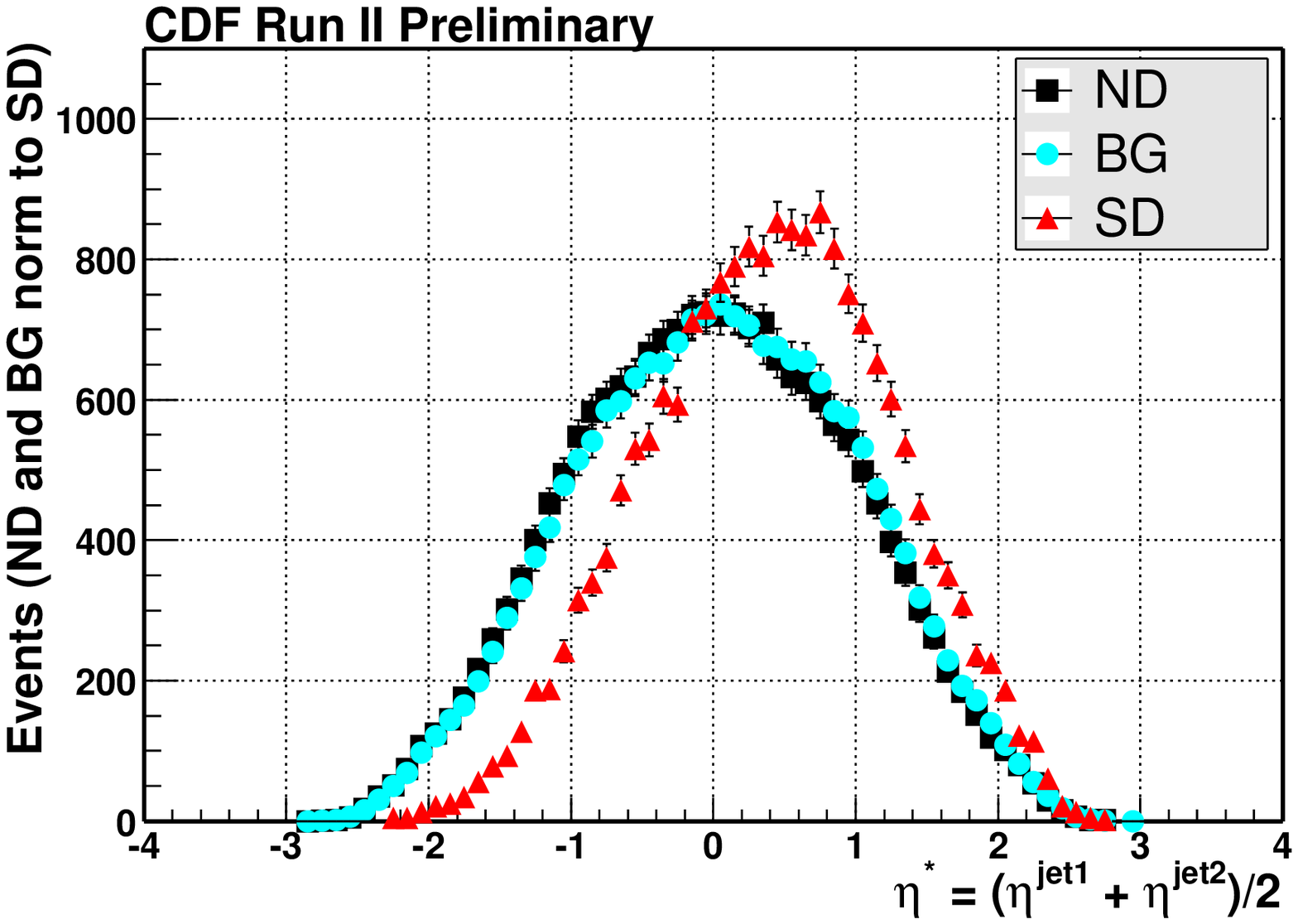,width=0.55\hsize}}
\caption{\label{Xi} 
{\em Left}: Momentum loss of the antiproton ($\xi_{\overline{p}}$) distribution in the RP+J5 and J5 samples.
SD and BG regions are selected according to the measured $\xi$ values;
{\em Right}: Average rapidity of the two leading jets.}
\end{figure}

In Run~II, a dedicated trigger (RP+J5) selects events with a three-fold RP coincidence and at least one 
calorimeter tower with $E_T>5$~GeV.
A further offline selection requires at least two jets of $E_T^{corr}>5$~GeV and $|\eta|<2.5$.
Jets are corrected for detector effects and underlying event corrections.
Calorimeter information alone is used to determine $\xi_{\overline{p}}=\frac{1}{\sqrt{s}}\sum_{i=1}^nE_T^ie^{-\eta^i}$, 
which is calculated using all calorimeter towers including MPs (Fig.~\ref{Xi}, left).
The declining of the distribution at $\xi_{\overline{p}}\sim 0.03$ occurs in the region where the RP acceptance is decreasing. 
A large number of events are at $\xi_{\overline{p}}\sim 1$, where the contribution comes from two sources:
diffractive dijets with a superimposed soft non-diffractive interaction, and non-diffractive dijets superimposed with a soft diffractive interaction.
The plateau observed between $0.02<\xi_{\overline{p}}<0.1$ (SD) results from a $d\sigma/d\xi\sim 1/\xi$ distribution, which is expected for diffractive production.
Mean dijet pseudorapidity distributions, $\eta^* = (\eta_1 + \eta_2) /2$, are shown in Figure~\ref{Xi} (right).
ND events are distributed around a mean value of $\eta^\star =0$, while SD events are shifted to positive $\eta^*$ values, 
indicating the boost of the center-of-mass of the interacting particles opposite to the recoil antiproton.

\begin{figure}[tp]
\epsfxsize=1.0\textwidth
\centerline{\epsfig{figure=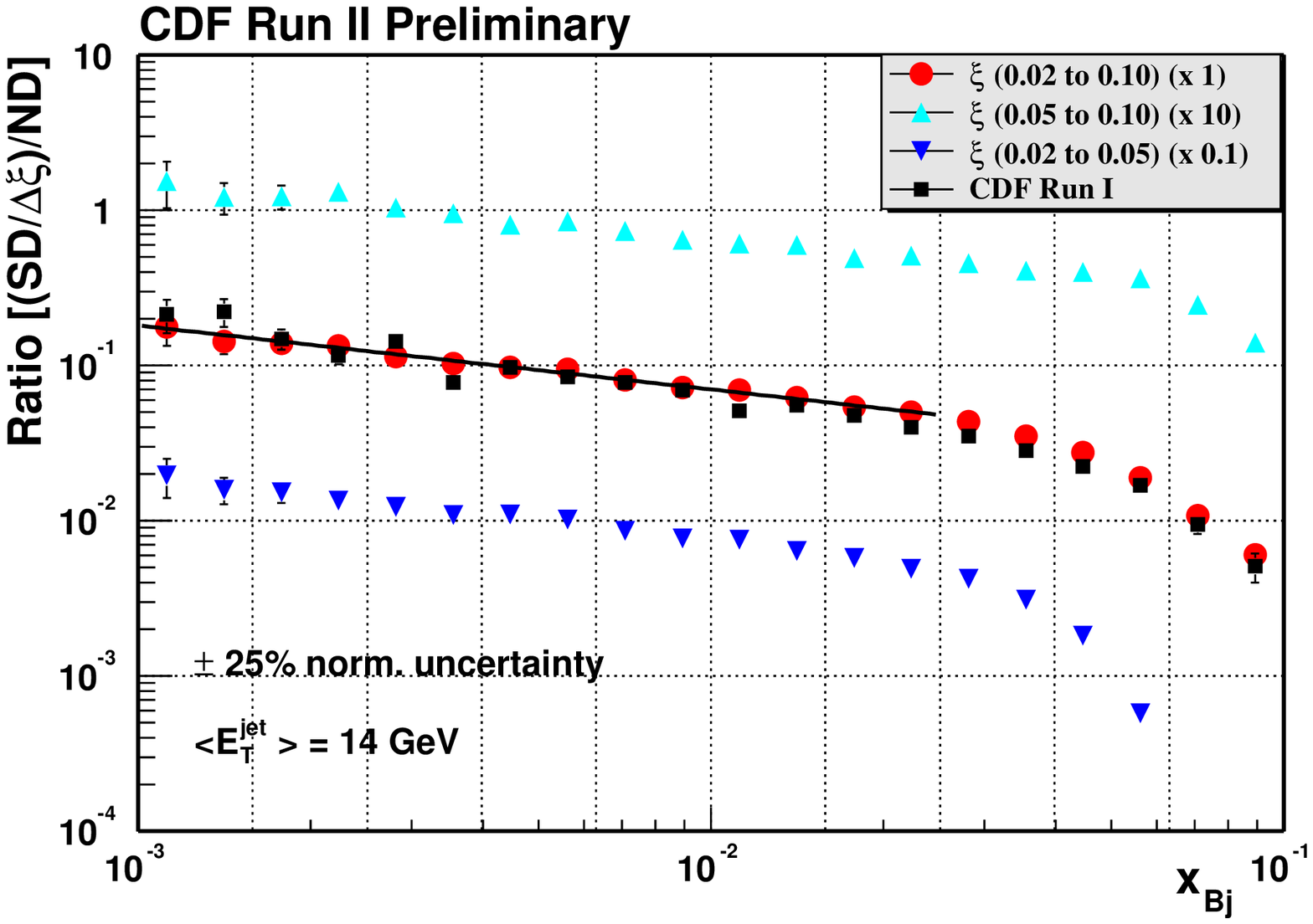,width=0.55\hsize}
\epsfig{figure=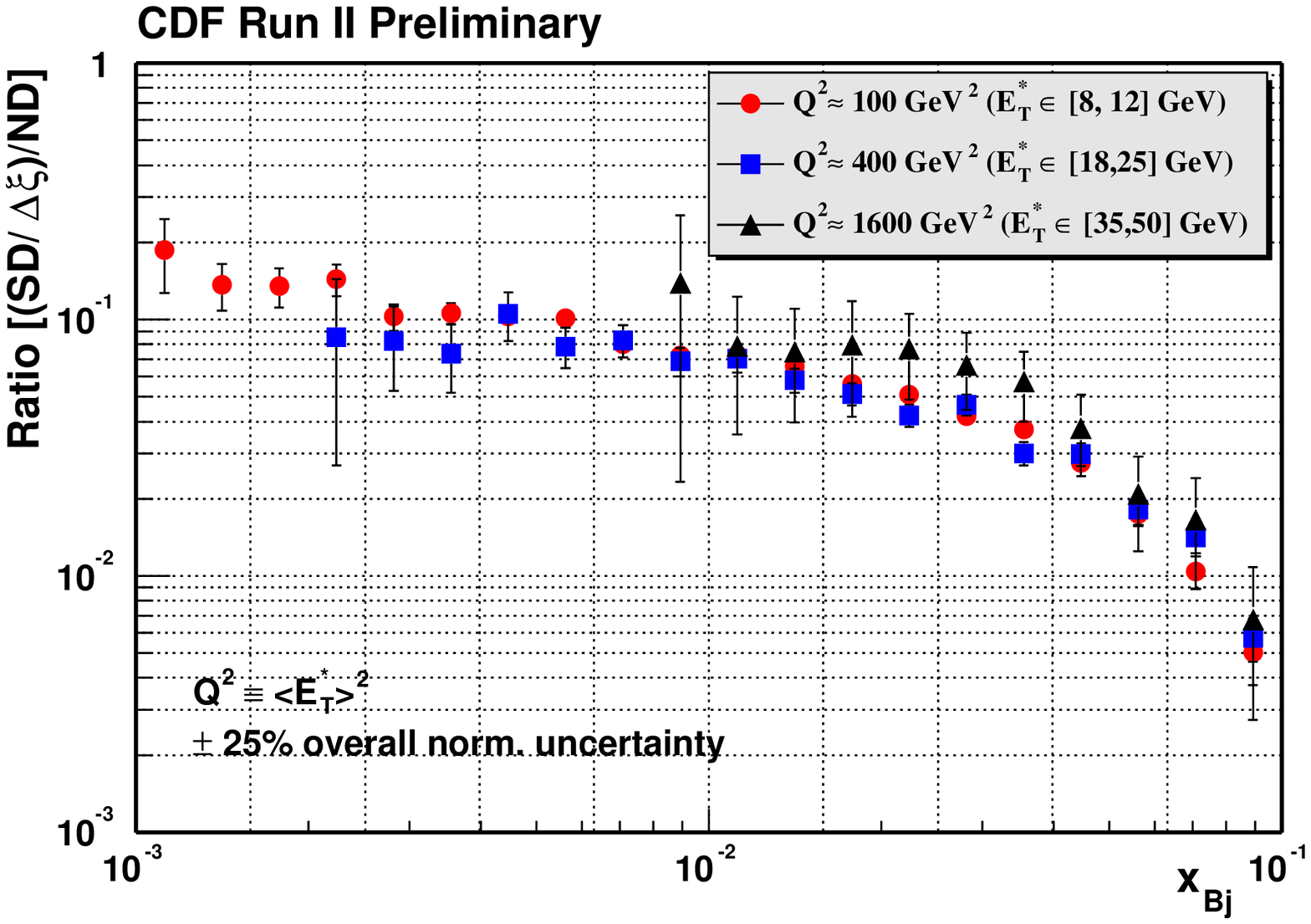,width=0.55\hsize}}
\caption{\label{run2_sf} 
Ratio of diffractive to non-diffractive dijet event rates as a function of $x_{Bj}$.
{\em Left}: Compared to Run~I; {\em Right}: For different values of $E{_T}^2 \equiv Q^2 $.}
\end{figure}

Measurement of the SD to ND event rate ratio is consistent with the Run~I result (Fig.~\ref{run2_sf}, left).
Furthermore, the jet $E_T$ spectrum extends to higher values than in Run~I.
A preliminary result indicates that the ratio does not depend strongly on $E{_T}^2 \equiv Q^2$
in the range from $Q^2 = 100$~GeV$^2$ up to 1600~GeV$^2$ (Fig.~\ref{run2_sf}, right).
The relative normalization uncertainty cancels out in the ratio.
This result indicates that the $Q^2$ evolution of the Pomeron is similar to that of the proton.

\section{Exclusive Dijet Production}

In Double Pomeron Exchange (DPE) dijet production,
a Pomeron is emitted from each nucleon and a Pomeron-Pomeron collision 
results in an exclusive dijet final state, produced together
with both the leading proton and anti-proton surviving the interaction and escaping in the very forward region.
At CDF, the RP can only detect the anti-proton.

\begin{figure}[tp]
\epsfxsize=1.0\textwidth
\centerline{
\epsfig{figure=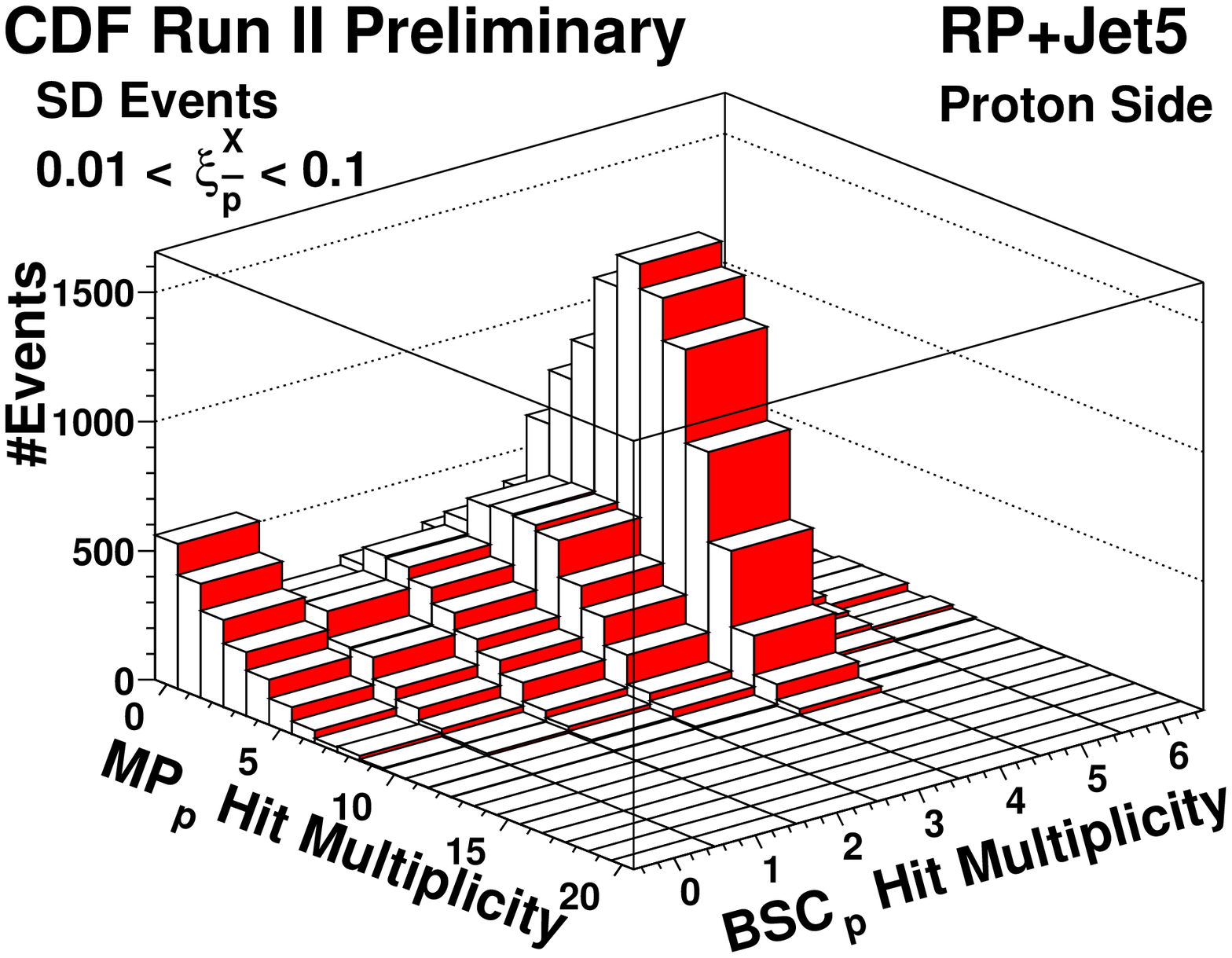,width=0.50\hsize}
\epsfig{figure=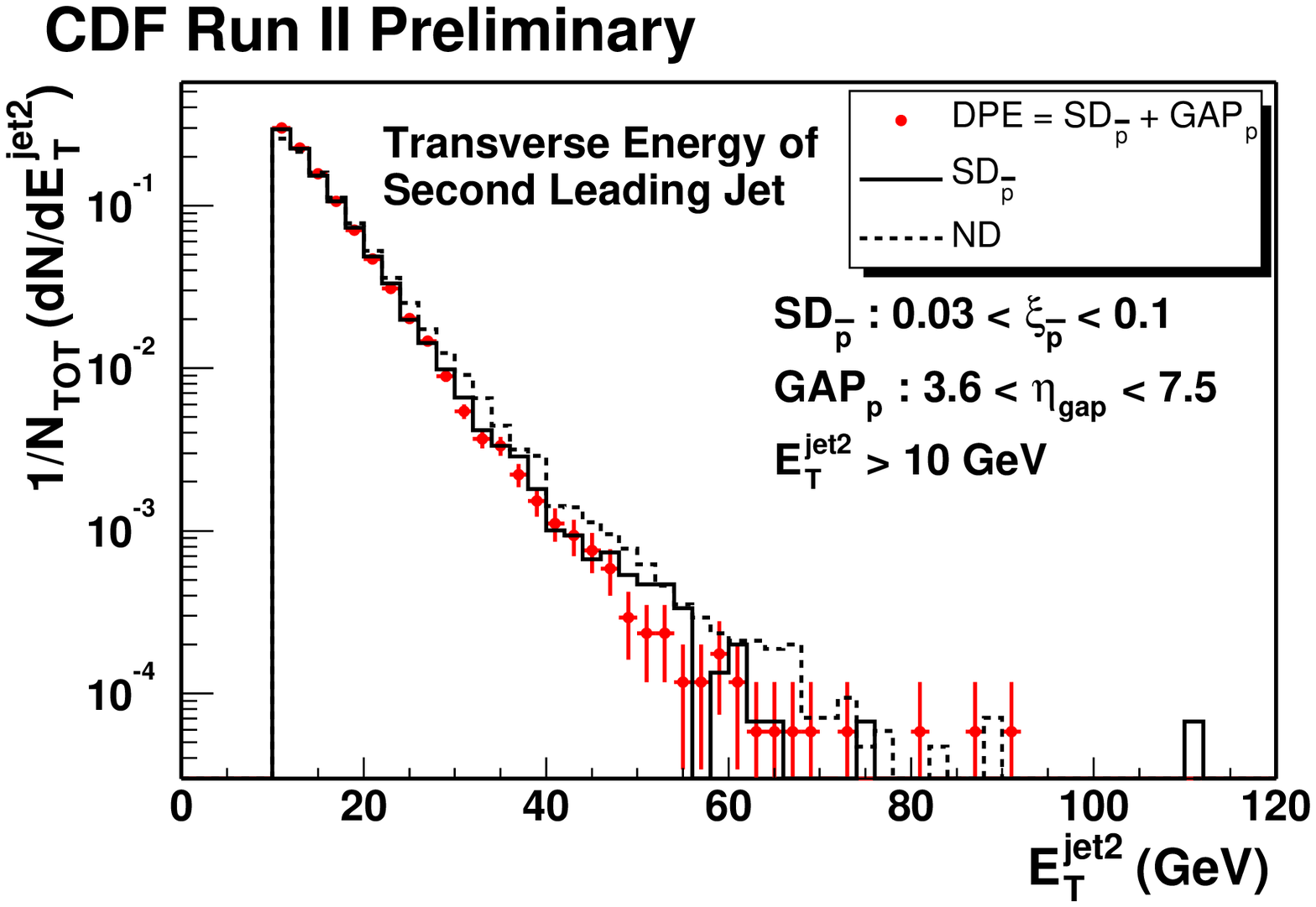,width=0.56\hsize}
}
\caption{\label{dpe_et_dphi}
{\em Left}: MP versus BSC multiplicity on the outgoing proton side in RP+J5 triggered events;
{\em Right}: Next-to-leading jet transverse energy distribution.
}
\end{figure}

\begin{figure}[thp]
\epsfxsize=1.0\textwidth
\centerline{
\epsfig{figure=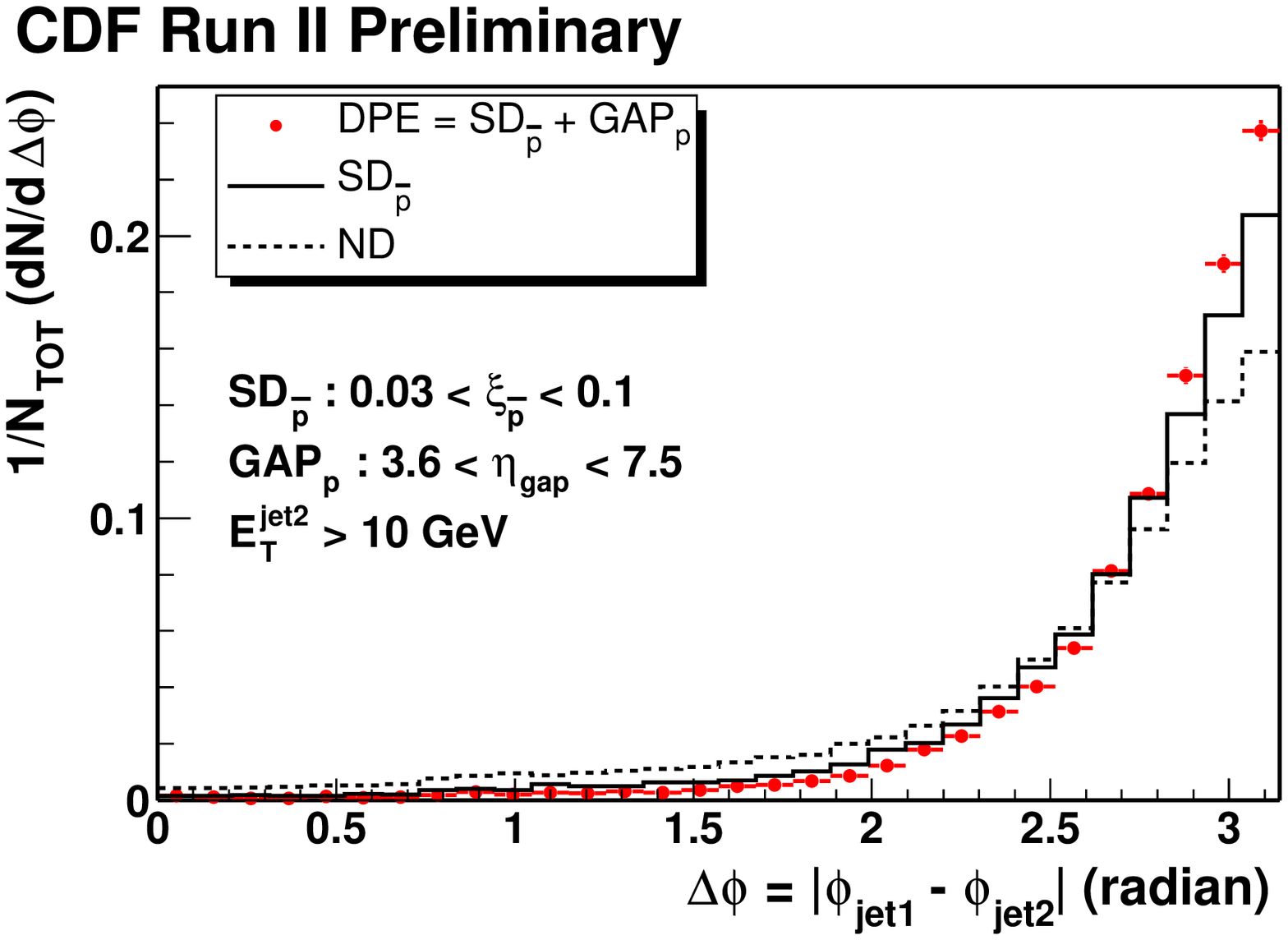,width=0.62\hsize}
\epsfig{figure=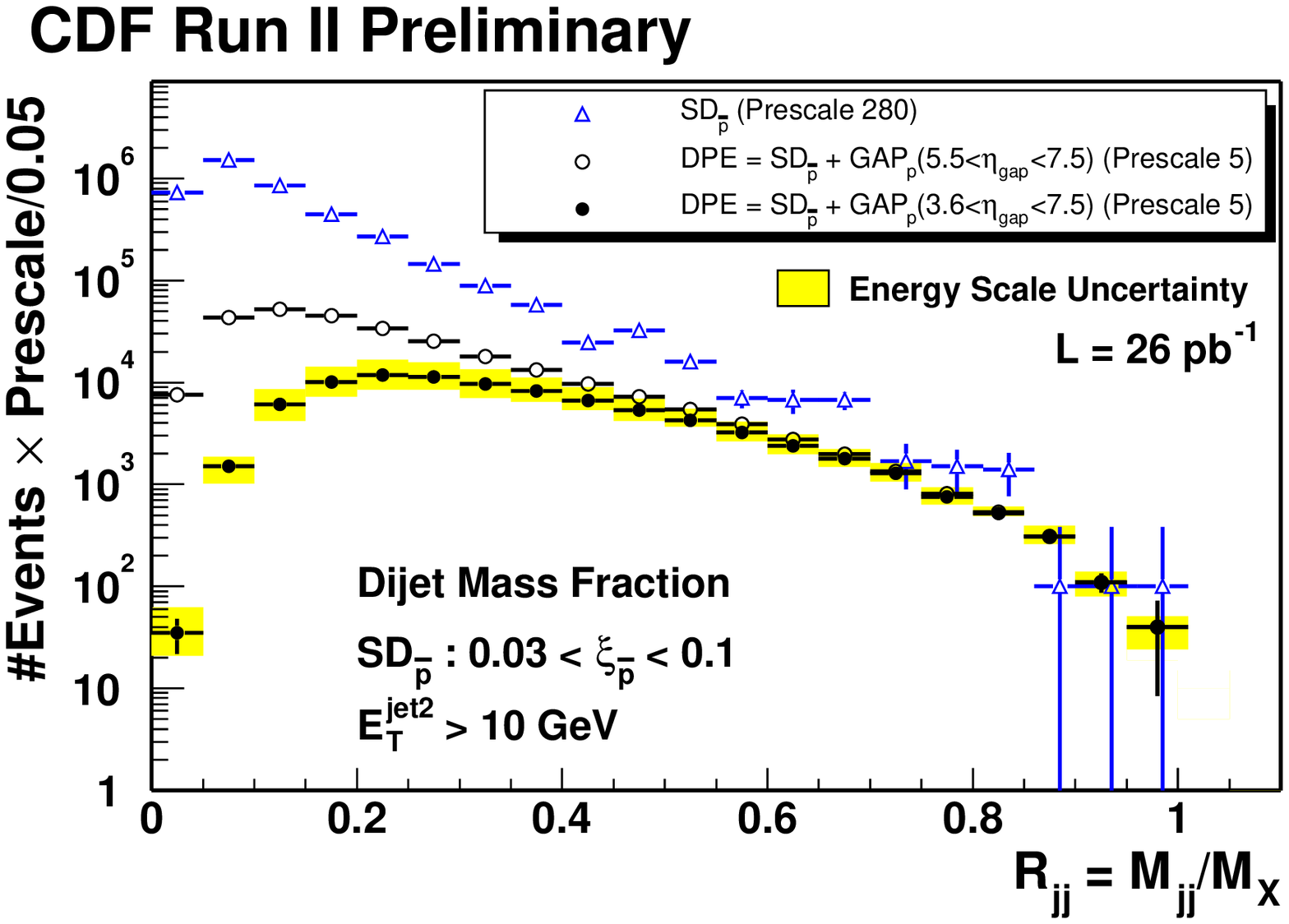,width=0.62\hsize}
}
\caption{\label{dpe_massratio}
{\em Left}:  Azimuthal angle difference between the two leading jets;
{\em Right}: Dijet mass fraction for different rapidity gap selections.
}
\end{figure}

In Run~II, the RP+J5 trigger, which was discussed in the previous section, can also be used to study DPE events.
This trigger selects a fraction of DPE events which can be isolated
by counting BSC and MP multiplicities on the proton side (Fig.~\ref{dpe_et_dphi}, left). 
The two peaks, at high and low multiplicity, are due to SD and DPE events, respectively.
Furthermore, a much larger sample has already been collected with a dedicated trigger.
The DPE trigger requires one RP coincidence, a proton-side rapidity gap in the BSCs, and at least one 
calorimeter tower ($E_T>$5~GeV) in the central detector.
Multiple interactions are rejected offline by requiring events with number of vertices $N_{vertex}\le 1$.
At least two jets ($E_T^{corr}>$10~GeV, $|\eta|<2.5$) are required.
The sample is further tightened by requiring the events to have $0.01<\xi_{\overline{p}}<0.1$. 
In order to reduce multiplicity fluctuations in SD events and thus enhance DPE events, 
a rapidity gap of $\sim 4$ units, including MP and BSC ($3.6<|\eta|<7.5$), is also required on the proton side.
In $\sim 26~pb^{-1}$ of data, the final sample consists of approximately 17,000 events.
The $E_T$ distributions for both leading and next-to-leading (Fig.~\ref{dpe_et_dphi}, right) jets 
are similar for ND, SD and DPE samples.
The dijet azimuthal angle difference shows that jets are more back-to-back in DPE events than in SD events 
(Fig.~\ref{dpe_massratio}, left).

The exclusive dijet production rate of these DPE events is of great interest to determine
the exclusive Higgs production cross section and prepare for the LHC experiments~\cite{higgs}.
The dijet mass fraction ($R_{jj}$), defined as the dijet invariant mass ($M_{jj}$) divided by the mass of the entire system 
$M_X =\sqrt{\xi_{\overline{p}}\cdot\xi_p\cdot s}$, is calculated by using all available energy in the calorimeter.
If jets are produced exclusively, $R_{jj}$ should be equal to one.
Uncorrected energies are used in Figure~\ref{dpe_massratio} (right) and no visible excess is evident over a smooth distribution.
After including systematic uncertainties, an upper limit on the exclusive dijet production cross section is calculated
based on events with $R_{jj}>0.8$ (Table~\ref{tab:xs}).

\begin{table}[h]
\begin{center}
\begin{tabular}{c|c}
\hline
minimum leading jet $E_T$ & cross section limit \\ \hline
10 GeV & $970\pm65(stat)\pm272(syst)$~pb \\
25 GeV & $34\pm5(stat)\pm10(syst)$~pb \\ \hline
\end{tabular}
\caption{\label{tab:xs} Exclusive dijet production cross section limit for events at $R_{jj}>0.8$.}
\end{center}
\end{table}

\section{Exclusive $J/\Psi+\gamma$ Production}

\begin{figure}[tp]
\epsfxsize=1.0\textwidth
\epsfig{figure=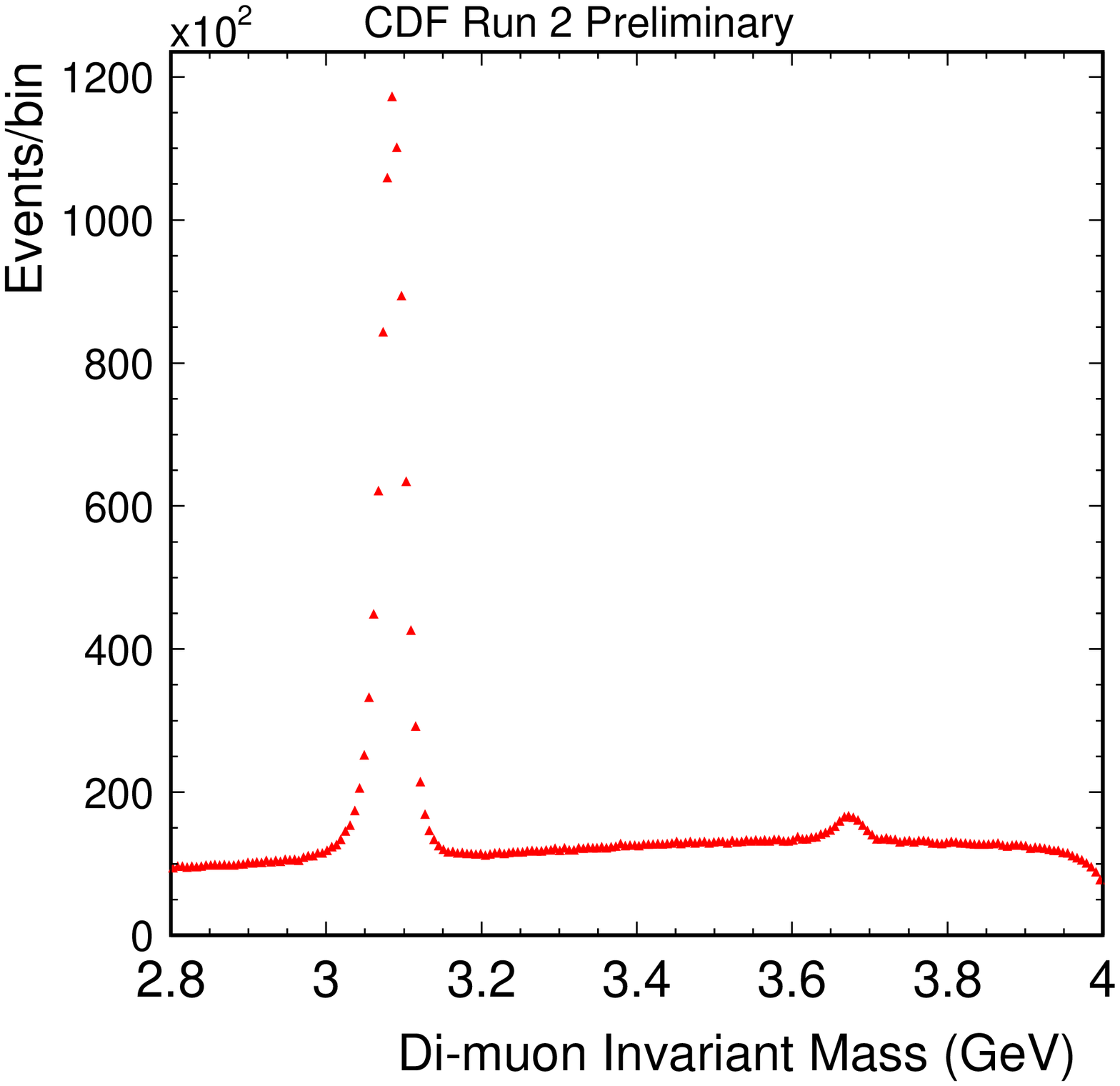,width=0.50\hsize}
\epsfig{figure=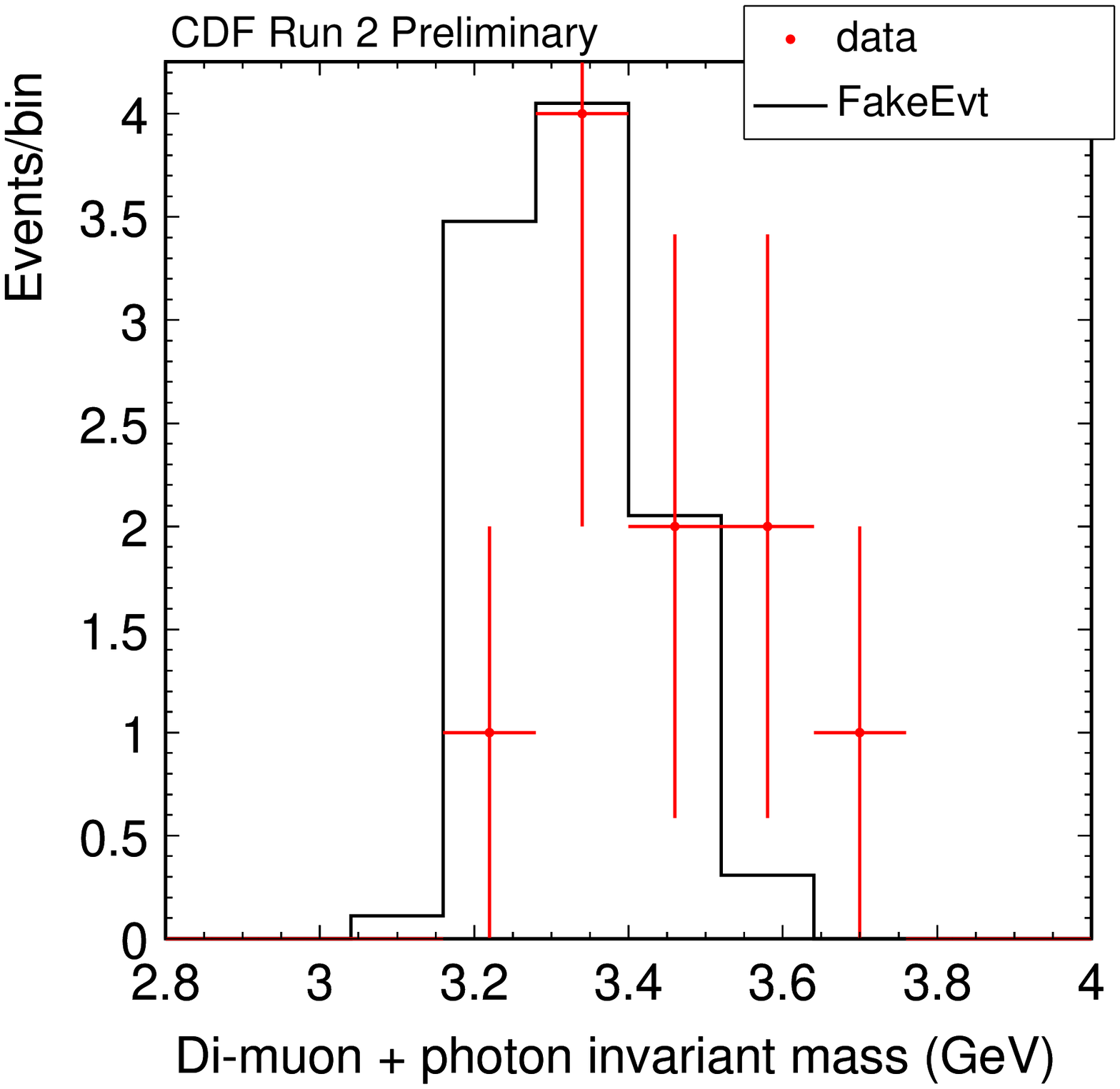,width=0.50\hsize}
\caption{\label{excl_jpsi}
{\em Left}: Invariant mass of the dimuon sample.
{\em Right}: Invariant mass of the muon pair plus the EM tower in the exclusive sample compared to Monte Carlo predictions.
}
\end{figure}

A process similar to the exclusive Higgs production is the exclusive production of $\chi^0_c$, as it has the 
same quantum numbers as the Higgs boson.
A cross section of $\sim 600$~nb is predicted at the Tevatron~\cite{khoze_chi}.
Using 93~pb$^{-1}$ of data collected with a di-muon trigger, $J/\Psi\rightarrow\mu^+\mu^-$ events are initially 
selected (Fig.~\ref{excl_jpsi}, left).
Calorimeter towers above threshold and reconstructed tracks are used to classify event multiplicities.
Events with hits in either the BSCs or MPs are rejected.
The Time-Of-Flight (TOF) detector is used to reject a potentially large source of cosmic ray background.
A total of 23 events are found with a muon pair in the $J/\Psi$ mass window, ten of which also contain an 
electromagnetic~(EM) calorimeter tower above threshold (Fig.~\ref{excl_jpsi}, right).
In the final sample, cosmic ray and fake 
``photon''~\cite{photon_quote} background sources are estimated to be negligible.
Multiplicity fluctuations due to calorimeter noise are expected to be small. However, it is experimentally difficult to 
evaluate their final contribution to the background, given the small number of events.
Therefore, the 10 events found are to be considered as an upper limit on exclusive production cross section.
After calculating trigger efficiency, detector and selection cut acceptances,
under the assumption that all observed ``dimuon plus EM tower'' candidates are from exclusive $J/\Psi+\gamma$ events, 
the resulting cross section is $\sigma(p\overline{p}\rightarrow p+J/\Psi+\gamma+\overline{p})=58\pm18(stat)\pm39(syst)$~pb.

\section{Conclusions}

Improved CDF forward detectors add new capabilities to an extended understanding of diffractive phenomena during Run~II.
A measurement of the diffractive structure function confirms Run~I results and indicates 
that the process is $Q^2$-independent, within current uncertainties.
Exclusive production of dijet and $\chi^0_c$ events is not found in the data and stringent cross section limits are set.
These results have been obtained during the first year of data-taking with well-performing detectors, and 
additional data are becoming available for further diffractive studies.


\begin{thebibliography}{99}
\bibitem{photon_quote} The name ``photon'' is in quotations as no quality identification cut is applied.
\end{thebibliography}
\end{document}